# MultiSoundGen: Video-to-Audio Generation for Multi-Event Scenarios via SlowFast Contrastive Audio-Visual Pretraining and Direct Preference Optimization

**Jianxuan Yang[1*†], Xiaoran Yang[1,2*], Lipan Zhang[1], Xinyue Guo[1], Zhao Wang[1], Gongping Huang[2]**

[1]MiLM Plus, Xiaomi Inc., China

[2]School of Electronic Information, Wuhan University, Wuhan, China

## Abstract

Current video-to-audio (V2A) methods struggle in complex multi-event scenarios (video scenarios involving multiple sound sources, sound events, or transitions) due to two critical limitations. First, existing methods face challenges in precisely aligning intricate semantic information together with rapid dynamic features. Second, foundational training lacks quantitative preference optimization for semantic-temporal alignment and audio quality. As a result, it fails to enhance integrated generation quality in cluttered multi-event scenes. To address these core limitations, this study proposes a novel V2A framework: MultiSoundGen. It introduces direct preference optimization (DPO) into the V2A domain, leveraging audio-visual pretraining (AVP) to enhance performance in complex multi-event scenarios. Our contributions include two key innovations: the first is SlowFast Contrastive AVP (SF-CAVP), a pioneering AVP model with a unified dual-stream architecture. SF-CAVP explicitly aligns core semantic representations and rapid dynamic features of audio-visual data to handle multi-event complexity; second, we integrate the DPO method into V2A task and propose AVP-Ranked Preference Optimization (AVP-RPO). It uses SF-CAVP as a reward model to quantify and prioritize critical semantic-temporal matches while enhancing audio quality. Experiments demonstrate that MultiSoundGen achieves state-of-the-art (SOTA) performance in multi-event scenarios, delivering comprehensive gains across distribution matching, audio quality, semantic alignment, and temporal synchronization. Demos are available at https://v2aresearch.github.io/MultiSoundGen/.

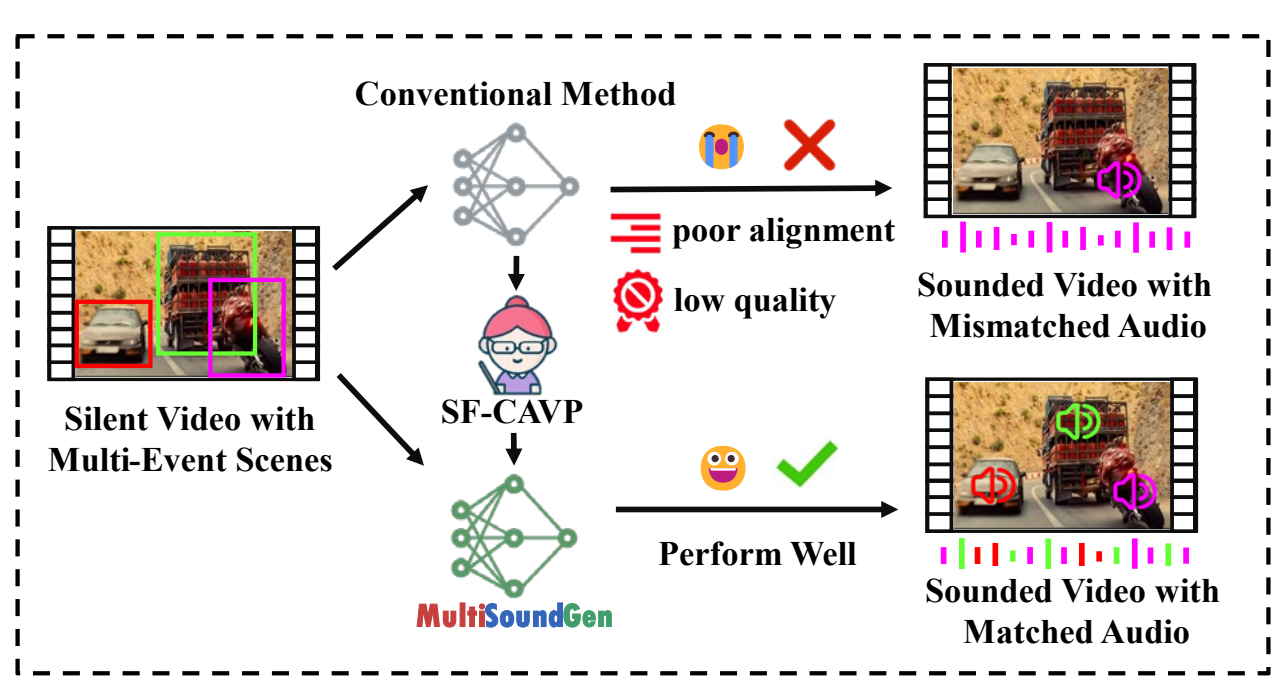

Figure 1. MultiSoundGen: a novel V2A framework for multi-event scenarios. It pioneers the integration of direct preference optimization into the V2A domain, leveraging audio-visual pretraining to improve both audio-visual alignment and audio quality in multi-event scenarios.

## Introduction

With the rapid development of artificial intelligence, video generation models have advanced significantly (Liu et al. 2024; Kong et al. 2024; Polyak et al. 2024). Models like Sora, Veo, and MovieGen can generate videos from text or images, but their synchronous sound effect generation remains subpar, often leaving videos silent. Video-to-audio (V2A) generation techniques, aiming to address this gap, have become a pressing research challenge, crucial for enhancing video quality and immersion (Xing et al. 2024; Wang et al. 2024; Cheng et al. 2025).

Existing V2A techniques work adequately in simple scenarios. However, in real-world settings with dense events, numerous elements, and frequent transitions, they struggle. Aligning intricate semantics with rapid dynamics is highly challenging. Moreover, foundational training lacks quantitative preference optimization for semantic-temporal alignment and audio quality, leading to semantic loss, semantic mismatch, poor synchronization, and degraded quality.

Direct preference optimization (DPO) emerges as a transformative solution by bypassing these limitations. Unlike traditional methods, DPO leverages preference signals to directly optimize the model's understanding of "high-quality generation" (Rafailov et al. 2023). This capability is irreplaceable by existing training frameworks. DPO's efficacy in related domains like text-to-audio (TTA) sets a precedent. For example, TangoFlux (Hung et al. 2024) uses language-audio pre-training (LAP) models to rank generated audio by text alignment, achieving SOTA fidelity by refining implicit text-audio associations. This raises the question: Can audio-video-pretraining (AVP)-based DPO enhance V2A performance in complex scenarios? However, applying it to V2A faces three challenges. First, no mature DPO solutions exist for V2A, necessitating new explorations. Second, aligning audio-video modalities is more complex than text-audio alignment due to strict temporal synchronization and semantic alignment requirements. Third, optimizing modality alignment must not compromise audio quality, demanding a balanced approach across multiple metrics.

*Equal contribution.

†Corresponding Author. Email: yangjianxuan@xiaomi.com

These challenges reduce to two core topics: designing effective AVP models and relevant preference optimization strategies. While image-text (e.g., CLIP) and audio-text (e.g., CLAP) pre-training have advanced (Radford et al. 2021; Wu et al. 2023), audio-video pre-training lags due to high synchronization demands and feature complexity.

To tackle these issues, we introduce SlowFast contrastive audio-video pretraining (SF-CAVP), the first AVP model with a unified audio-video encoding architecture, tailored for multi-event scenarios. Its dual-stream SlowFast design, slow paths capturing core semantics, fast paths tracking rapid dynamics (Feichtenhofer et al. 2019; Xiao et al. 2020; Kazakos et al. 2021), adapts to dense sound events and transitions. Integrating SlowFast features across audio and video aligns cross-modal features of multiple sound sources, ensuring robust semantic alignment and temporal synchronization in complex scenarios. Complementing SF-CAVP, we propose AVP-Ranked Preference Optimization (AVP-RPO). It uses SF-CAVP as a reward model to rank the base model's generated audio by similarity to the input video, then constructs preference pairs to align the base model via DPO.

Together, these form MultiSoundGen (Figure 1), a novel V2A model for complex multi-event scenarios. Comparisons experiments show it outperforms SOTA baselines in distribution matching, audio quality, semantic alignment, and temporal synchronization—with up to 10.3% improvement in distribution matching and 5.3% in temporal synchronization for multi-event videos. It also performs robustly in single-event and out-of-distribution scenarios, securing SOTA in distribution matching and audio-visual alignment. In summary, our key contributions are as follows:

- We develop SF-CAVP, the first AVP model with a unified SlowFast architecture, aligning both core semantics and rapid dynamics of audio-visual features.
- We introduce AVP-RPO, a method extending DPO to the V2A domain. It leverages SF-CAVP to enhance both audio-video alignment and audio quality of the base model.
- Our MultiSoundGen V2A model achieves SOTA performance with comprehensive improvements in multi-event scenarios. It also maintains robust performance in both single-event and out-of-distribution general scenarios.

## Related Works

**Video-to-Audio Generation.** V2A generation aims to synthesize audio that is semantically aligned and temporally synchronized with video content. Early methods like SpecVQGAN (Iashin and Rahtu 2021) and Im2Wav (Sheffer and Adi 2023) explored autoregressive audio token generation from visual features, while VTA-LDM (Xu et al. 2024) leveraged latent diffusion with CLIP-based visual encoders to enhance semantic relevance. V2A-Mapper (Wang et al. 2024) maps CLIP visual embeddings to CLAP's audio-text space to guide generation. FoleyCrafter (Zhang et al. 2024) leverages CLIP-derived visual features with ControlNet for temporal coherence. These approaches all rely on text-mediated pretraining frameworks: CLIP-like methods learn visual features by aligning them with text semantics, while CLAP-like ones derive audio representations through associations with textual descriptions. However, text lacks the high-precision temporal cues critical for V2A, especially in complex multi-event scenarios—where precise frame-level synchronization is essential. This limitation underscores the need for AVP, which directly models intrinsic audio-visual correlations without textual mediation.

**Audio-Video Pretraining.** AVP methods are mainly categorized into two types: feature fusion (Gong et al. 2022; Liu et al. 2025) and contrastive learning (Sun et al. 2024; Jawade et al. 2025; Li et al. 2025; Tsiamas et al. 2025) approaches. Due to the effectiveness in capturing cross-modal relationships, contrastive learning-based AVP methods are preferred in multi-modal generation tasks like V2A.

Segment AVCLIP (Iashin et al. 2024), a contrastive-AVP (CAVP) approach, aligns audio-visual features by encoding audio with AST (Gong et al. 2021) and video with Motionformer (Patrick et al. 2021). MMAudio (Cheng et al. 2025) uses the MM-DiT (Esser et al. 2024) architecture with conditional synchronization, leveraging Segment AVCLIP to generate visual synchronization features for generation guidance. V-AURA (Viertola, Iashin and Rahtu 2024), an autoregressive V2A model, relies on Segment AVCLIP for temporal alignment of visual and auditory features. Diff-Foley (Luo et al. 2023) proposes a V2A method based on Latent Diffusion Model (LDM), using CAVP—with audio encoded by PANNs (Kong et al. 2020) and video by SlowOnly (Feichtenhofer et al. 2019)—to learn aligned features .

In summary, existing AVP methods often select separate audio/video encoders without considering multi-modal collaborative encoding. This paper first proposes an AVP model with a unified SlowFast dual-stream architecture for audio-video encoding, enhancing audio-visual alignment accuracy and cross-modal representational richness.

**Direct Preference Optimization.** In Large Language Models (LLM), DPO is widely used to leverage off-the-shelf reward models to enhance performance (Ouyang et al. 2022). For TTA, TangoFlux introduces CRPO, which uses CLAP to improve alignment and quality of audio generation. To the best of our knowledge, our method is the first work to introduce DPO into V2A.

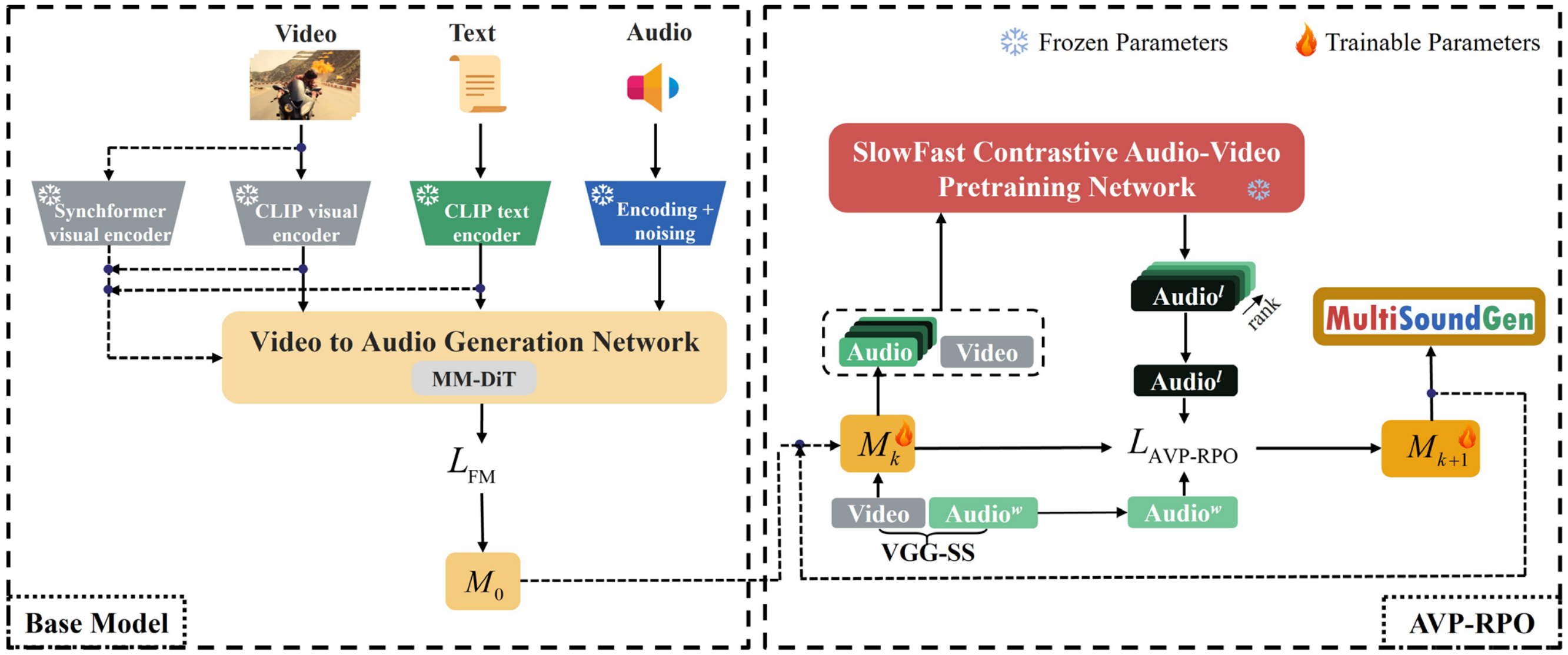

Figure 2. Overview of MultiSoundGen network. The backbone of MultiSoundGen is MM-DiT trained with a CFM objective. Two key innovations underpin MultiSoundGen: SF-CAVP and AVP-RPO. AVP-RPO uses SF-CAVP as a reward model to iteratively optimize the base model, boosting audio-video alignment and audio quality.

## Method

This paper presents MultiSoundGen, a novel V2A method specialized in complex multi-event video scenarios powered by AVP-based DPO alignment. As shown in Figure 2, the backbone of MultiSoundGen is multimodal diffusion transformer (MM-DiT) trained with a conditional flow matching (CFM) objective (Tong et al. 2024). Two key innovations underpin MultiSoundGen: first, the SF-CAVP, the first AVP model with a unified SlowFast dual-stream architecture, tailored to address V2A with multi-event scenario. Second, the AVP-RPO method. It uses SF-CAVP as a reward model to directly optimize the base model, boosting audio-video alignment and audio quality. This section will elaborates on SF-CAVP and AVP-RPO. Introductions to the MM-DiT architecture and CFM strategy are provided in Appendix A.

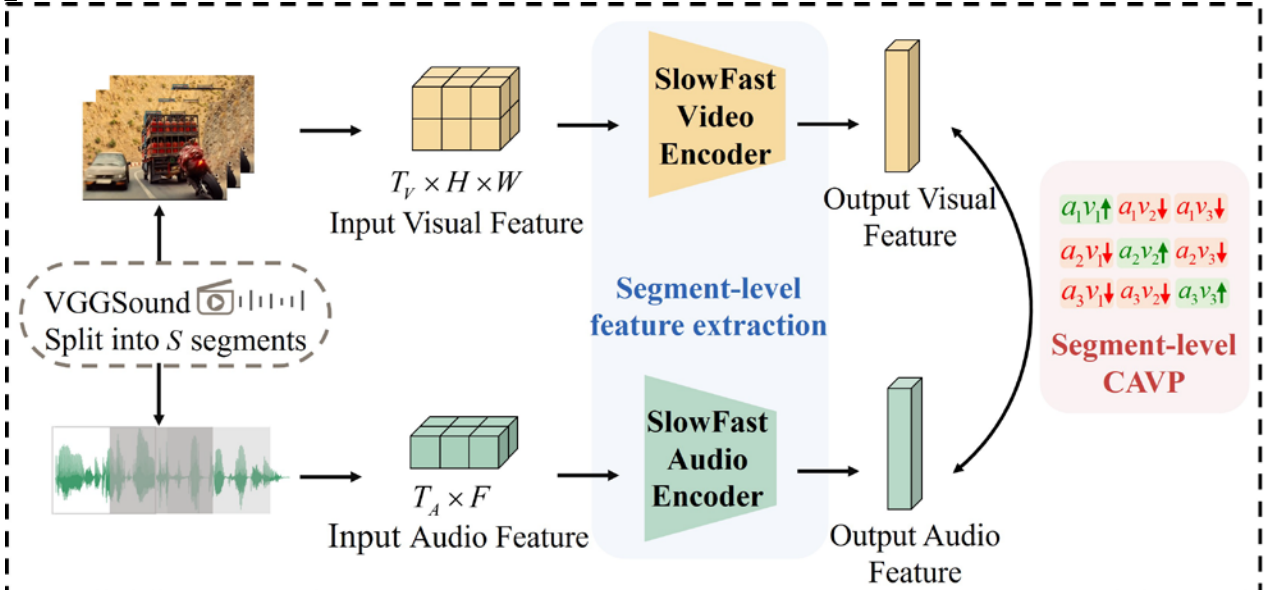

Figure 3. SF-CAVP framework. Features are extracted from temporal segments of the video by SlowFast video encoders and SlowFast audio encoders. Then, segment-level contrastive pre-training is performed.

### SF-CAVP

Figure 3 illustrates the framework of SF-CAVP. In this pre-training method, features are extracted from temporal segments of the video by SlowFast video encoders and SlowFast audio encoders. Then, segment-level contrastive pre-training is performed. These key steps will be elaborated on in the following content. For further details and implementation of this step, readers are referred to Appendix B.

**Segment-level SlowFast feature extraction.** The visual streams and the paired audio are split into $S$ segments with equal duration. Segment-level input size of the video encoder is $T_V \times H \times W$, where $T_V$ is the temporal length of the video segment, $H$ is the height , and $W$ is the width; Segment-level input size of the audio encoder is $T_A \times F$, where $T_A$ and $F$ are the temporal and frequency length of the log-mel-spectrogram respectively.

For each segment $s \in \{1,...,S\}$, we extract audio and visual features using their respective encoders. One of the highlights of this study lies in the high uniformity of the audio and video encoder architectures: they adopt the same SlowFast structure and share identical key parameters. For both encoders, the slow stream has a lower sampling rate, which is $1/\alpha$ times that of the fast stream, while its channel capacity is higher, being $\beta$ times that of the fast stream. The SlowFast architectures employ multi-level lateral connections to fuse features from the fast to the slow stream across stages.

**Segment-level CAVP.** In CAVP, the key elements are as follows:

First, the definition of positive and negative pairs. A positive pair refers to a pair of audio and visual segments extracted from the same time interval of the same video. In contrast, negative pairs consist of segments from the same video (but different time intervals) and segments from other videos within the batch. Second, the contrastive loss function. We adopt the InfoNCE loss (Oord, Li and Vinyals 2018) derived by averaging two directional losses. Specifically, $L_{av}$ is calculated as

$$L_{av} = -\frac{1}{BS}\sum_{i=1}^{BS}\log\frac{\exp(a_i \cdot v_i / \tau)}{\sum_{j=1}^{BS}\exp(a_i \cdot v_j / \tau)}, \tag{1}$$

where $B$ is the batch size, $S$ is the segments number, $a_i, i \in \{1,...,S\}$ is the segment-level audio feature, $v_i, v_j$ are the segment-level visual features, and $\tau$ stands for a trainable temperature parameter. The counterpart loss $L_{va}$ is defined analogously, thus the total loss is defined as:

$$L = (L_{av} + L_{va})/2 \tag{2}$$

**AVP-RPO**

AVP-RPO uses SF-CAVP as a reward model to rank the base model's generated audios according to their similarity with the input video. Then, preference pairs are built to align the base model. First, we take the SOTA V2A model, MMAudio, as the base model for alignment, denoted as $M_0$. After that, AVP-RPO iteratively aligns $M_k$ to $M_{k+1}$, $k \in [0, N_{\text{it}} - 1]$, where $N_{\text{it}}$ denotes the total number of iterations. Each alignment iteration has three steps: a) batched data generation, b) SF-CAVP-based ranking and preference creation, and c) fine-tuning $M_k$ to $M_{k+1}$ via DPO.

**Batched Data Generation.** In iteration $k$, for each video in the training set, the model $M_k$ generates $N_a$ different audio clips.

**SF-CAVP-based ranking and preference creation.** SF-CAVP is employed to rank the similarity between the generated audio and the input video. For each audio clip $A_k, k \in [1, N_a]$, processing the audio and video with SF-CAVP yields $S$ audio-visual features pairs, thus we can get $S$ cosine similarities. There are two key improvements in this step which better guide the following DPO procedure: a) The final similarity score $s_{\text{fs}}$ is calculated using the following order statistics formula instead of the global average:

$$s_{\text{fs}} = \text{mean}(s_{\text{sim}(1)}, ..., s_{\text{sim}(\lfloor S/4 \rfloor)}), \tag{3}$$

where $s_{\text{sim}(i)}$ denotes the $i$th smallest cosine similarity among all $S$ cosine similarities. Such calculation enhances the discriminability in score ranking. b) Unlike CRPO which regards the audio with the highest score as the winner, we adopt the input video's ground truth audio as the winner. This is because, unlike text-audio pairs, audio and video have a strictly one-to-one correspondence. Thus, the ground truth audio is more appropriate as the winner than the highest-scoring generated audio. The loser remains the lowest-scoring generated audio.

**DPO fine-tuning.** DPO has proven effective in imbuing LLM with human preferences, enabling them to generate outputs that better align with human expectations. This approach has been successfully adapted to diffusion models through DPO-Diffusion, which facilitates alignment in such generative models (Wallace et al. 2023). Building on the work of Esser et al. (2024), DPO-Diffusion loss function can be applicable to CFM, replacing noise-matching loss terms in DPO-Diffusion with flow-matching terms:

$$\begin{aligned} L_{\text{DPO-FM}} = -\mathbb{E}_{t,x_t^w,x_t^l,\mathbf{C}} \log\sigma(-\beta_w(&\underbrace{\| v_\theta(t,\mathbf{C},x_t^w) - u_t^w \|^2}_{\text{winning loss}} \\ -\underbrace{\| v_\theta(t,\mathbf{C},x_t^l) - u_t^l \|^2}_{\text{losing loss}} - (&\underbrace{\| v_{\theta_{\text{ref}}}(t,\mathbf{C},x_t^w) - u_t^w \|^2}_{\text{winning reference loss}} \\ -\underbrace{\| v_{\theta_{\text{ref}}}(t,\mathbf{C},x_t^l) - u_t^l \|^2}_{\text{losing reference loss}}&))). \end{aligned} \tag{4}$$

Here, $x_t^w$ and $u_t^w$ stand for the winning audio sample and flow velocity at timestep $t$. $x_t^l$ and $u_t^l$ represent the losing audio sample and flow velocity, respectively. As shown in Eq. (4), $L_{\text{DPO-FM}}$ centers on the relative likelihood of winning and losing responses. It minimizes loss by widening the gap between winning and losing losses, even as both losses rise.

To address this issue, we integrate the DPO loss with the flow matching loss of the winning sample, namely, $L_{\text{FM-win}} = \mathbb{E}_{t,x_t^w,\mathbf{C}} \left\| v_\theta\left(t,\mathbf{C},x_t^w\right) - u_t^w \right\|^2$, into the optimization objective:

$$L_{\text{AVP-RPO}} = N(L_{\text{DPO-FM}}) + N(L_{\text{FM-win}}). \tag{5}$$

$N(\cdot)$ means normalizing the loss term to $[0,1]$. This integration not only enhances preference ranking but also anchors the model in learning the attributes of high-quality data, thereby avoiding distortions in alignment and fidelity caused by over-optimization. In Eq. (5), the two loss terms are first normalized separately and then summed. The normalization step prevents either loss term from dominating gradient updates.

To address performance degradation from full fine-tuning, we use a "freeze bottom layers + optimize top layers" strategy to preserve the MM-DiT's diffusion and denoising mechanisms. We freeze underlying multimodal transformer blocks and earlier single-modal blocks—critical for fundamental cross-modal representations and generative flow. Instead, we fine-tune the last single-modal transformer layer, along with subsequent adaptive layer normalization (adaLN) layers (Perez et al. 2018) and 1D-convolution layers. The last single-modal layer refines final audio latents while maintaining diffusion dynamics. AdaLN layers (for frame-level temporal alignment via token modulation) and 1D-Conv layers (which capture local temporal structures) are optimized to refine local details without compromising the global denoising mechanisms.

| | Method | Params | Distribution matching | | | Audio Quality | Sem. Align. | Temp. Synch. |
|---|---|---|---|---|---|---|---|---|
| | | | $FD_{VGG}$ ↓ | $FD_{PANNs}$ ↓ | $KL_{PANNs}$ ↓ | IS ↑ | IB-score ↑ | Desync ↓ |
| Multi-Event Generation | Foleycrafter | 1.22B | 4.023 | 47.174 | 1.819 | 6.181 | 0.297 | 1.243 |
| | V-AURA | 695M | 4.340 | 44.572 | 1.748 | 5.284 | 0.312 | 0.797 |
| | Seeing&Hearing | 415M | 7.585 | 64.009 | 2.595 | 3.416 | **0.374** | 1.289 |
| | V2A-Mapper | 229M | **2.915** | 42.959 | 2.258 | 5.519 | 0.246 | 1.260 |
| | Frieren | 159M | 4.099 | 40.196 | 1.802 | 5.647 | 0.273 | 0.775 |
| | MMAudio | 157M | 4.100 | 39.519 | 1.256 | 6.295 | 0.342 | 0.379 |
| | MultiSoundGen | 157M | 3.716 | **39.073** | **1.244** | **6.354** | 0.343 | **0.360** |
| Single-Event Generation | Foleycrafter | 1.22B | 4.664 | 31.530 | 1.996 | 9.510 | 0.292 | 1.311 |
| | V-AURA | 695M | 3.739 | 29.930 | 1.874 | 8.144 | 0.305 | 0.782 |
| | Seeing&Hearing | 415M | 5.911 | 49.057 | 2.922 | 4.872 | **0.382** | 1.296 |
| | V2A-Mapper | 229M | 1.828 | 25.270 | 2.325 | 8.668 | 0.252 | 1.204 |
| | Frieren | 159M | 2.715 | 30.353 | 2.432 | 8.312 | 0.245 | 0.902 |
| | MMAudio | 157M | **1.701** | **22.495** | 1.421 | **9.695** | 0.321 | 0.407 |
| | MultiSoundGen | 157M | 1.751 | 22.623 | **1.420** | 9.640 | 0.322 | **0.399** |
| Kling-Audio-Eval | Foleycrafter | 1.22B | **2.352** | 16.304 | 2.606 | 7.130 | **0.284** | 1.211 |
| | V-AURA | 695M | 7.193 | 29.694 | 3.235 | 6.940 | 0.261 | 0.887 |
| | Frieren | 159M | 4.321 | 22.880 | 3.482 | 6.489 | 0.183 | 1.177 |
| | MMAudio | 157M | 4.321 | 9.755 | 2.464 | **7.362** | 0.275 | **0.575** |
| | MultiSoundGen | 157M | 4.253 | **9.513** | **2.433** | 7.277 | 0.272 | 0.577 |

Table 1. V2A results of multi-event generation, single-event generation and Kling-Audio-Eval benchmark. The first and second places are bolded and underlined, respectively. MultiSoundGen achieves SOTA performance with improvement across distribution matching, audio quality, semantic alignment, and temporal synchronization compared to the base model. Notably, the framework also maintains robust performance in both single-event and out-of-distribution general scenarios.

## Experiments

### Dataset

**SF-CAVP training.** We use VGGSound (Chen et al. 2020) to train SF-CAVP. VGGSound contains around 200k 10-second YouTube video clips across 310 audio classes. Following the original train splits, we conducted initial filtering to exclude empty or excessively short videos, resulting in a training set of 173K videos and a validation set of 2K videos.

**AVP-RPO training.** We train AVP-RPO on the VGG-Sound Source (VGG-SS) dataset (Chen et al. 2021), which is derived from VGGSound test set. VGG-SS has a high-quality bounding box annotations for sounding objects with 5K clips across 220 categories. Given that VGG-SS lacks a formally defined separation of training and testing data, we randomly select 4.4K pairs from the dataset for model training, 120 pairs for validation and 500 pairs for testing.

**Evaluation.** To validate the generation capability of MultiSoundGen, we manually split the VGG-SS test set into two subsets: VGG-SS-single (VGG-SS-S), comprising 356 samples of simple single sound events, and VGG-SS-multi (VGG-SS-M), consisting of 144 samples targeting complex multi-event scenarios. To demonstrate MultiSoundGen's generalization capability, we evaluate it on the out-of-distribution benchmark Kling-Audio-Eval (Wang et al. 2025). Kling-Audio-Eval is the first industrial-grade multimodal benchmark containing 21K human-annotated samples.

### Metrics

This paper evaluates the generation performance from four aspects: distribution matching, audio quality, semantic alignment, and temporal synchronization.

**Distribution matching.** This metric assesses how well the feature distribution of the generated audio matches the ground-truth audio. We use the Fréchet Distance (FD) and Kullback–Leibler (KL) distance (Wang et al. 2024). FD is computed via embeddings of PANNs (Kong et al. 2020) and VGGish (Gemmeke et al. 2017). Notice that PANNs generate global features, while VGGish processes short 0.96-second clips. The KL distance is determined using PANNs.

**Audio quality.** This metric assesses the audio quality of the generated audio via the inception score (IS). The IS is calculated using PANNs (Wang et al. 2024).

**Semantic alignment.** This metric assesses the similarity between the generated audio and the video. This score is the average cosine similarity between visual features of the video and audio features of the generated audio. Both features are extracted using ImageBind (Girdhar et al. 2023).

**Temporal synchronization.** This metric assesses how well the audio and video are synchronized in time with the DeSync score from Synchformer (Iashin et al. 2024). This score indicates the misalignment in seconds.

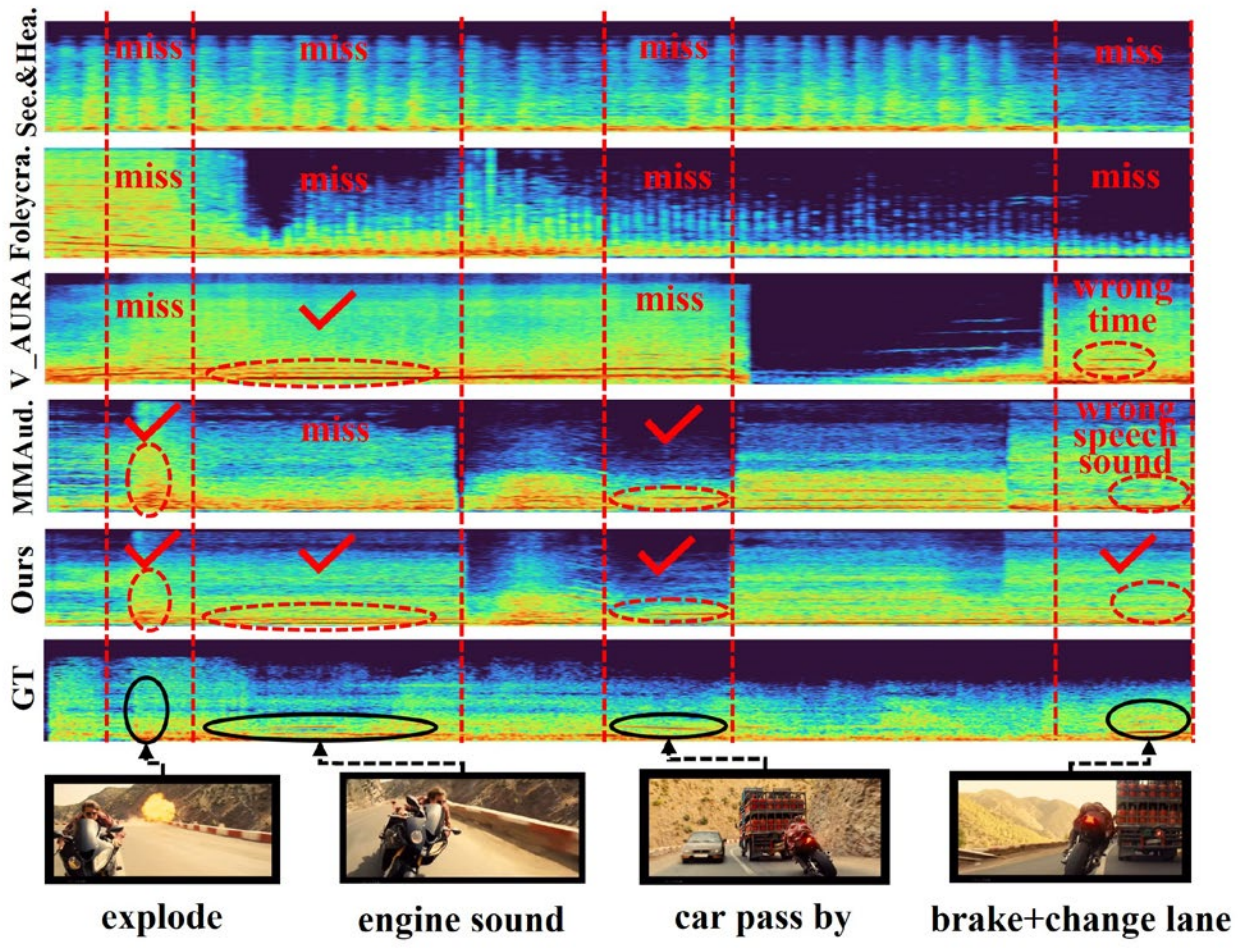

Figure 4. V2A result for a typical multi-event scenario. MultiSoundGen achieves robust audio-visual alignment.

## Main result

Table. 1 compares the main results of our MultiSoundGen with the base model, MMAudio (Cheng et al. 2025), and other competitive models, namely, Frieren (Wang et al. 2024), FoleyCrafter (Zhang et al. 2024), V-AURA (Viertola et al. 2024), Seeing and Hearing (Xing et al. 2024) and V2A-Mapper (Wang et al. 2024). Note that, to ensure the fairness of the comparison and demonstrate the effectiveness of the AVP-RPO method, MultiSoundGen is identical to the base model in all parameter settings except for the optimized model weight parameters. To observe the performance saturation phenomenon of DPO, optimization iteration $N_{it}$ is set to 5. We use results of the first iteration for comparison, with the reasons explained in the following content. All implementation details and descriptions of comparative methods are in Appendix C. First, we assess the results of multi-event generation. With just 4.4K DPO training data, MultiSoundGen outperforms MMAudio on all metrics with up to 10.3% improvement in distribution matching and 5.3% in temporal synchronization. This is enabled by SF-CAVP's multi-dimensional alignment capability and AVP-RPO's ability to quantify and prioritize audio-visual alignment and audio quality. MultiSoundGen also attains SOTA metrics in distribution matching, audio quality, semantic alignment, and temporal synchronization. Note that Seeing-and-Hearing directly optimizes the IB-score during the denoising process (Xing et al. 2024), so it gets the highest IB-score. This is consistent with the phenomenon observed by Cheng et al. (2025). For the single-event generation, MultiSoundGen ranks top 2 across all metrics. For the Kling-Audio-Eval benchmark, MultiSoundGen also ranks top 2 across all metrics expect for semantic alignment. We also present a generation result for a typical video of complex multi-event scenario: a motorcycle chase clip from a film. As shown in Figure 4, the video contains multiple sound events and frequent shot switches. While comparative methods suffer from semantic loss, semantic mismatch, and poor temporal synchronization, MultiSoundGen captures key sound events and achieves robust audio-visual alignment.

Figure 5 shows the variation trend of model performance with all five optimization iterations. "Iteration = 0" represents the base model. All subsequent experiments are denoted in the same way. Notably, IS continuously rises as iterations proceed, and $FD_{VGG}$ generally shows an optimizing trend. However, DeSync starts to fluctuate after the first iteration, while IB-score begins to fluctuate after the second iteration. This also reflects the difficulty of optimizing audio-video alignment while simultaneously taking into account the optimization of other important metrics. It can be concluded that, in the early stages of AVP-RPO iteration, all metrics are optimized. Subsequently, the timings of performance saturation for different metrics differ, which is related to factors such as the preference of the reward model.

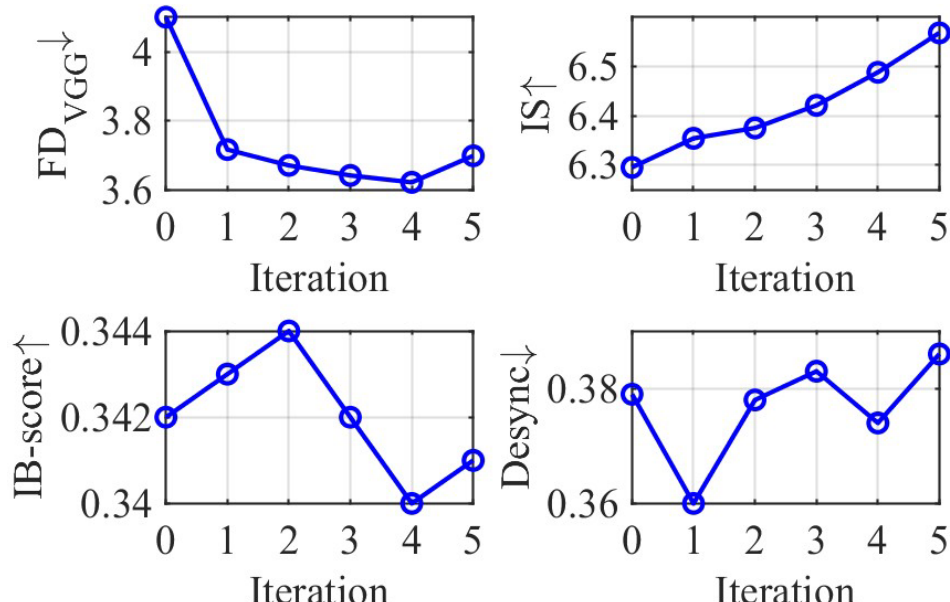

Figure 5. Model performance over five optimization iterations. Results indicate that timings of performance saturation for different metrics differ.

## Ablations

For all ablation experiments, we evaluate distribution matching ( $FD_{VGG}$ ), audio quality (IS), semantic alignment (IB-score), and temporal synchronization (DeSync) on the VGG-SS-M test set. To observe the phenomenon of permance saturation, we compare the results from all five iterations of AVP-RPO. Key ablation experiments are listed below, other ablation experiments will be in Appendix D.

**AVP for AVP-RPO.** To validate the effectiveness of SF-CAVP as the reward model in AVP-RPO, we compare it with another AVP method: Segment AVCLIP. Experimental results for the two AVP methods are provided in the

Figure 6. Using Segment AVCLIP as the reward model in AVP-RPO leads to an overall degradation of model performance, and the performance becomes increasingly poor with iterations. The experimental results indicate that AVP directly affects the performance of AVP-RPO, and also verify the effectiveness of SF-CAVP as a reward model.

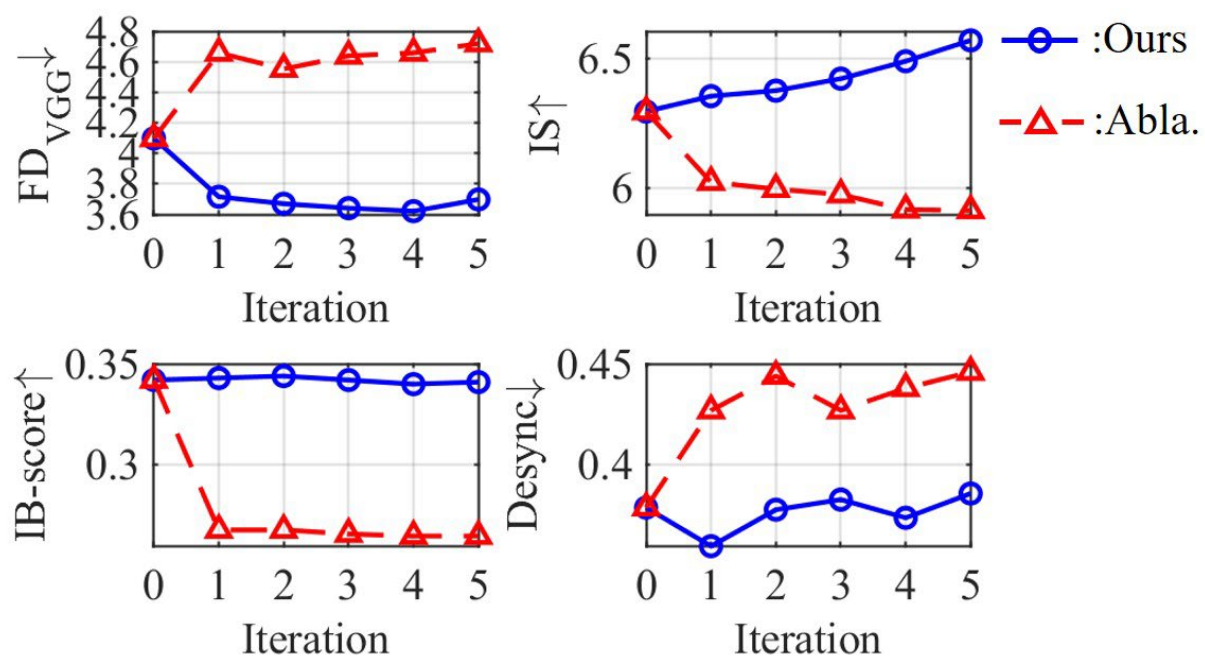

Figure 6. Ablation study on AVP selection in AVP-RPO. Segment AVCLIP (dashed line) leads to overall performance degradation (worsening with iterations), validating SF-CAVP's effectiveness as a reward model in AVP-RPO.

**Loss function for AVP-RPO.** We evaluate whether integrating $L_{\mathrm{FM-win}}$ into the optimization objective can reduce alignment and fidelity distortions from over-optimization. This is done by optimizations using $L_{\mathrm{AVP\text{-}RPO}}$ and $L_{\mathrm{DPO\text{-}FM}}$ (which excludes $L_{\mathrm{FM-win}}$) respectively. Results are shown in Figure 7. When using $L_{\mathrm{DPO\text{-}FM}}$, the overall optimization performance is largely consistent with that of using $L_{\mathrm{AVP\text{-}RPO}}$ but inferior. When MultiSoundGen adopts $L_{\mathrm{AVP\text{-}RPO}}$ as the optimization objective, all metrics outperform those of the former, with only Desync remaining on par.

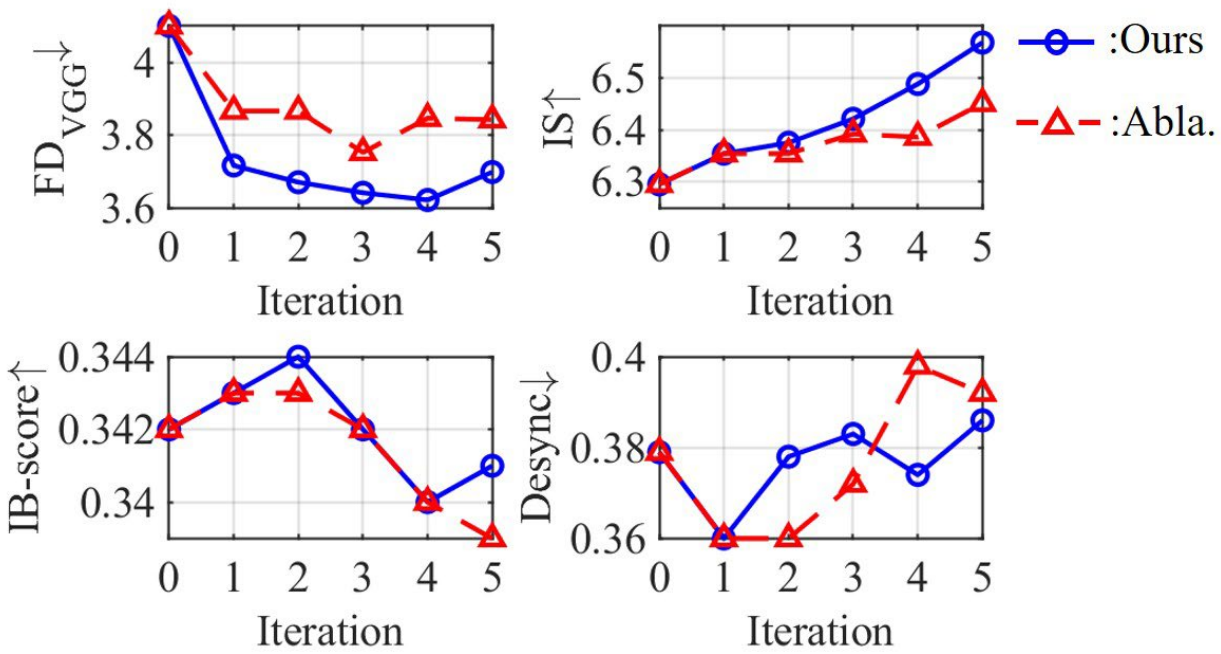

Figure 7. Comparison of optimization objectives. Method with $L_{\mathrm{DPO\text{-}FM}}$ (dashed line) shows optimization performance, but worse than that with $L_{\mathrm{AVP\text{-}RPO}}$ (solid line).

**Fine-tuning strategy.** To determine if full fine-tuning of the base model degrades performance, we compare it with our "freezing bottom layers + optimizing top layers" strategy. Experimental outcomes for both approaches are provided in the Figure 8. Full fine-tuning causes severe performance degradation. This arises from disrupting the MM-DiT model's diffusion process and denoising mechanism, leading to significant noise in generated audio.

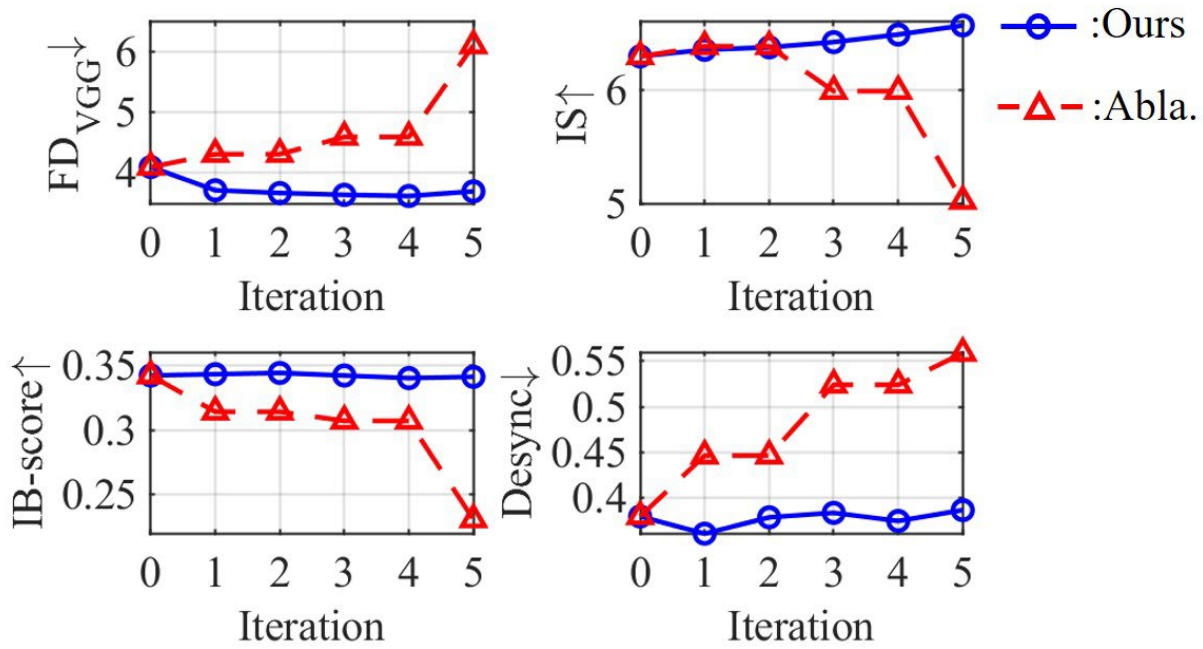

Figure 8. Comparison of full fine-tuning (dashed line) vs. "freezing bottom layers + optimizing top layers" (solid line). Full fine-tuning causes severe performance degradation, while the latter strategy performs much better.

## Limitation and Future Work

While our method has made notable progress, it has limitations. Constrained by the SlowFast framework, the SF-CAVP model only conducts contrastive learning on pooled features. Referencing methods like HiCMAE (Sun et al. 2014), CSMP (Li et al. 2025), and SCAV (Tsiamas et al. 2025) to implement more fine-grained contrastive learning could further enhance audio-video alignment accuracy. Additionally, AVP-RPO's training dataset is small-scale with room for improvement in quality. Future work will use larger, higher-quality data for optimization, potentially yielding better results, especially for large models.

## Conclusion

This paper presents MultiSoundGen, a novel V2A generation framework specifically designed for complex multi-event scenarios. Our key innovation lies in adapting DPO to the V2A domain, coupled with the newly proposed SF-CAVP module that serves as a reward model. This integrated approach enables comprehensive optimization of the base model in multi-event settings. Experiments demonstrate that MultiSoundGen achieves state-of-the-art performance with improvement across distribution matching, audio quality, semantic alignment, and temporal synchronization compared to the base model. Notably, the framework also maintains robust performance in both single-event and out-of-distribution general scenarios.

## Appendix A. MM-DiT architecture and CFM strategy

**A.1. MM-DiT.** MM-DiT (Esser et al. 2024) is a transformer architecture built to handle multiple modalities. It processes modalities as one-dimensional tokens and uses joint attention for cross-modal communication. MM-DiT features modality-specific weight streams to preserve each modality's distinct characteristics while enabling information flow. Aligned positional embeddings and convolutional MLPs enhance temporal alignment and local structure modeling across modalities.

**Basic Architecture and Inputs**. MM-DiT is built upon the DiT architecture (Hoogeboom, Heek, and Salimans. 2023) and operates in the latent space of a pretrained autoencoder when training models. Similar to other approaches, MM-DiT encodes the conditioning using pretrained models. Meanwhile, it constructs a sequence of embeddings from the inputs.

**Multimodal Processing Mechanism**. Due to the conceptual differences between different embeddings, the MM-DiT architecture uses separate sets of weights to handle different modalities. This is equivalent to having independent transformers for each modality, but the sequences of modalities are joined during the attention operation. This allows both representations to work in their own spaces while considering each other, facilitating information flow between token streams.

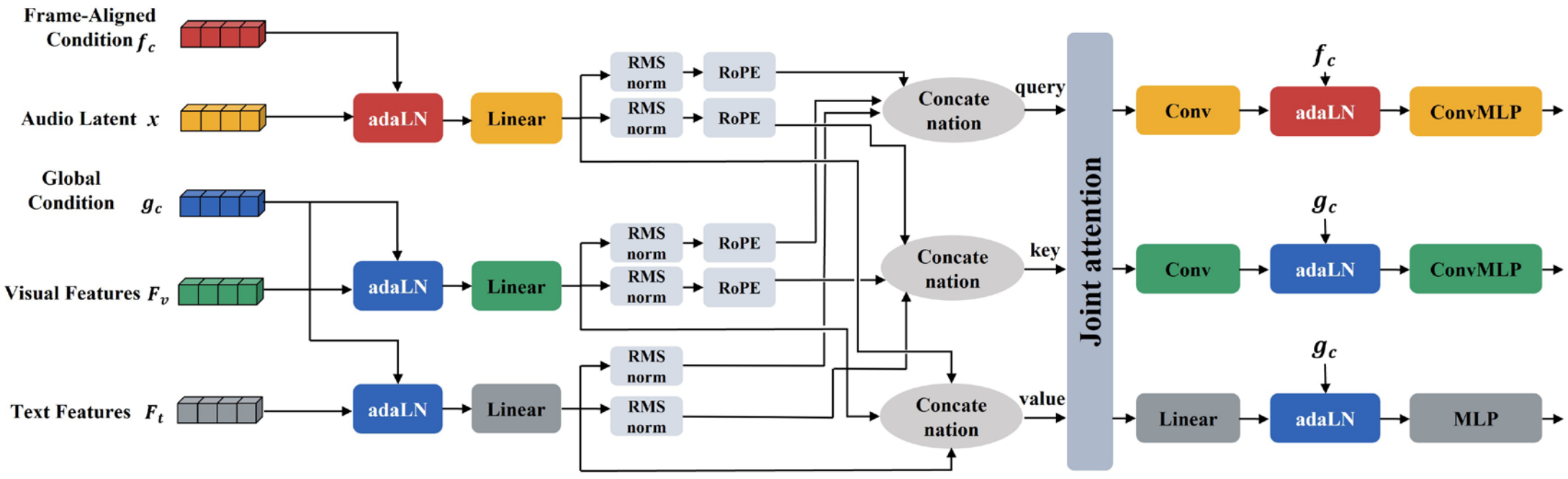

Figure A1 The architecture of MM-DiT.

As shown in Figure A1, the model's implementation processes three key modalities: audio, visual, and text. Audio is handled by encoding mel-spectrograms into latent vectors $x$ using a pretrained Variational Autoencoder (VAE) at a frame rate of 31.25 fps. Visual features $F_v$ are processed by two branches: a CLIP encoder providing 1024-dimensional features at 8 fps and a Synchformer visual encoder supplying 768-dimensional synchronization features at a higher 24 fps for precise alignment. Text features $F_t$, consisting of 77 tokens, each 1024-dimensional, are also extracted via a CLIP encoder. To guide the generation process, MM-DiT incorporates two distinct conditioning mechanisms. A global condition $g_c$ is introduced via adaptive layer normalization (adaLN) layers by combining the flow time step with average-pooled visual and text features, creating a shared vector that is broadcast across all tokens. Additionally, a frame-aligned condition $f_c$ is used to improve temporal synchronization. This involves upsampling the high-frame-rate synchronization features from the Synchformer to match the 31.25 fps of the audio stream, then injecting this condition into the audio stream's adaLN layer for fine-grained, token-level control.

**A.2. CFM strategy**

Conditional flow matching (CFM) (Tong et al. 2024) is adopted as a training objective for generative modeling. For sampling at test time, noise $x_0$ is first drawn from the standard normal distribution. An ODE solver is employed to perform numerical integration from $t=0$ to $t=1$. This integration follows a learned conditional velocity vector field $v_\theta(t,\mathbf{C},x):[0,1]\times\mathbb{R}^C\times\mathbb{R}^d\to\mathbb{R}^d$, where $t$ is the timestep, $\mathbf{C}$ represents conditions, and $x$ is a point in the vector field. The velocity vector field is represented by a deep network with parameters $\theta$. During the training phase, we determine $\theta$ by taking the CFM objective into account:

$$\mathbb{E}_{t,q(x_0),q(x_1,\mathbf{C})}\left\|v_\theta\left(t,\mathbf{C},x_t\right)-u_t\right\|^2, \tag{A1}$$

where $t\in[0,1]$. Here, $q(x_0)$ is the standard normal distribution. $q(x_1,\mathbf{C})$ is sampled from the training data.

$$x_t = tx_1+\left(1-t\right)x_0 \tag{A2}$$

establishes a linear interpolation path between the noise and the data.

$$u_t = x_1 - x_0 \tag{A3}$$

signifies the corresponding flow velocity at $x_t$.

## Appendix B. SlowFast video encoder and audio encoder of SF-CAVP

One of the highlights of this study lies in the high uniformity of the audio and video encoder architectures: they adopt the same SlowFast structure and share identical key parameters. For both encoders, the slow stream has a lower sampling rate, which is $1/\alpha$ times that of the fast stream, while its channel capacity is higher, being $\beta$ times that of the fast stream. The parameters are set as $\alpha = 4, \beta = 8$. The SlowFast architectures employ multi-level lateral connections to fuse features from the fast to the slow stream across stages. The final representation is of length 2304 for both encoders.

| stage | *Slow* pathway | *Fast* pathway | Output sizes $T \times S^2$ |
|---|---|---|---|
| raw clip | - | - | $32 \times 224^2$ |
| data layer | stride $16, 1^2$ | stride $2, 1^2$ | $Slow: 8 \times 224^2$<br>$Fast: 32 \times 224^2$ |
| conv$_1$ | $1 \times 7^2, 64$<br>stride $1, 2^2$ | $5 \times 7^2, 8$<br>stride $1, 2^2$ | $Slow: 8 \times 112^2$<br>$Fast: 32 \times 112^2$ |
| pool$_1$ | $1 \times 3^2$ max<br>stride $1, 2^2$ | $1 \times 3^2$ max<br>stride $1, 2^2$ | $Slow: 8 \times 56^2$<br>$Fast: 32 \times 56^2$ |
| res$_2$ | $\begin{bmatrix} 1 \times 1^2, 64 \\ 1 \times 3^2, 64 \\ 1 \times 1^2, 256 \end{bmatrix} \times 3$ | $\begin{bmatrix} 1 \times 1^2, 8 \\ 1 \times 3^2, 8 \\ 1 \times 1^2, 32 \end{bmatrix} \times 3$ | $Slow: 8 \times 56^2$<br>$Fast: 32 \times 56^2$ |
| res$_3$ | $\begin{bmatrix} 1 \times 1^2, 128 \\ 1 \times 3^2, 128 \\ 1 \times 1^2, 512 \end{bmatrix} \times 4$ | $\begin{bmatrix} 3 \times 1^2, 16 \\ 1 \times 3^2, 16 \\ 1 \times 1^2, 64 \end{bmatrix} \times 4$ | $Slow: 8 \times 28^2$<br>$Fast: 32 \times 28^2$ |
| res$_4$ | $\begin{bmatrix} 3 \times 1^2, 256 \\ 1 \times 3^2, 256 \\ 1 \times 1^2, 1024 \end{bmatrix} \times 6$ | $\begin{bmatrix} 3 \times 1^2, 32 \\ 1 \times 3^2, 32 \\ 1 \times 1^2, 128 \end{bmatrix} \times 6$ | $Slow: 8 \times 14^2$<br>$Fast: 32 \times 14^2$ |
| res$_5$ | $\begin{bmatrix} 3 \times 1^2, 512 \\ 1 \times 3^2, 512 \\ 1 \times 1^2, 2048 \end{bmatrix} \times 3$ | $\begin{bmatrix} 3 \times 1^2, 64 \\ 1 \times 3^2, 64 \\ 1 \times 1^2, 256 \end{bmatrix} \times 3$ | $Slow: 8 \times 7^2$<br>$Fast: 32 \times 7^2$ |

Table B1. Network structure of the SlowFast video encoder. The backbone is ResNet-50. Strides are denoted by $\{S_T, S_S{}^2\}$. Here, $S_T$ is the temporal stride, $S_S{}^2$ is the spatial stride. Kernel dimensions are denoted by $\{T \times S^2, C\}$. Here, $T$ is the temporal size, $S^2$ is the spatial size, and $C$ is the channel size.

### B.1. Slowfast video encoder

As shown in Table B1, the implementation structure of the SlowFast video encoder (Feichtenhofer et al. 2019) is based on ResNet-50 (He et al. 2016) as the backbone. The complete model is formed through the specific design of the Slow pathway, Fast pathway, and lateral connections, with the detailed structure as follows:

**Slow Pathway. Input Sampling**: From a 32-frame raw video clip, $N_{vs}=8$ frames are sparsely sampled with a temporal stride $\tau=4$, i.e., 1 frame is taken every 4 frames. **Convolution Design**: It is a temporally strided 3D ResNet variant. Except for the early layers that use non-degenerate temporal convolutions (temporal kernel size > 1), the convolution kernels in subsequent layers are essentially 2D convolution kernels. No temporal downsampling is performed to avoid performance degradation when the input stride is large.

**Fast Pathway. Input Sampling**: Based on the same raw video clip as the Slow pathway, $\alpha N_{vs}=32$ frames are sampled with a temporal stride of $\tau/\alpha=1$. **Channels and Convolutions**: The channel capacity is $1/\beta$ of that of the Slow pathway. Each block uses non-degenerate temporal convolutions, and there are no temporal downsampling layers to maintain high temporal resolution features.

| stage | *Slow* pathway | *Fast* pathway | Output sizes $T\times F$ |
| --- | --- | --- | --- |
| spectrogram | - | - | $128\times128$ |
| data layer | stride $4,1$ | stride $1,1$ | $Slow: 8\times128$<br>$Fast: 32\times128$ |
| conv$_1$ | $1\times7, 64$<br>stride $2,2$ | $5\times7, 8$<br>stride $2,2$ | $Slow: 8\times64$<br>$Fast: 32\times64$ |
| pool$_1$ | $3\times3$ max<br>stride $2,2$ | $3\times3$ max<br>stride $2,2$ | $Slow: 8\times32$<br>$Fast: 32\times32$ |
| res$_2$ | $\begin{bmatrix}1\times1,64\\1\times3,64\\1\times1,256\end{bmatrix}\times3$ | $\begin{bmatrix}1\times1,8\\1\times3,8\\1\times1,32\end{bmatrix}\times3$ | $Slow: 8\times32$<br>$Fast: 32\times32$ |
| res$_3$ | $\begin{bmatrix}1\times1,128\\1\times3,128\\1\times1,512\end{bmatrix}\times4$ | $\begin{bmatrix}3\times1,16\\1\times3,16\\1\times1,64\end{bmatrix}\times4$ | $Slow: 8\times16$<br>$Fast: 32\times16$ |
| res$_4$ | $\begin{bmatrix}3\times1,256\\1\times3,256\\1\times1,1024\end{bmatrix}\times6$ | $\begin{bmatrix}3\times1,32\\1\times3,32\\1\times1,128\end{bmatrix}\times6$ | $Slow: 8\times8$<br>$Fast: 32\times8$ |
| res$_5$ | $\begin{bmatrix}3\times1,512\\1\times3,512\\1\times1,2048\end{bmatrix}\times3$ | $\begin{bmatrix}3\times1,64\\1\times3,64\\1\times1,256\end{bmatrix}\times3$ | $Slow: 8\times4$<br>$Fast: 32\times4$ |

Table B2. Network structure of the SlowFast audio encoder. The backbone is ResNet-50. Strides are denoted by $\{S_T, S_F\}$. Here, $S_T$ is the temporal stride, $S_F$ is the frequency stride. Kernel dimensions are denoted by $\{T\times F, C\}$. Here, $T$ is the temporal size, $F$ is the frequency size, and $C$ is the channel size.

**Lateral Connections. Connection Positions**: Lateral connections are set after each stage of ResNet (such as after $\text{pool}_1$, $\text{res}_2$, etc.) to fuse features from the Fast pathway into the Slow pathway. **Feature Matching**: Due to the different temporal dimensions of the two pathways, the lateral connections perform transformations to match the feature sizes, such as time-to-channel conversion, time-strided sampling, and time-strided convolution. The fusion methods can be element-wise summation or concatenation, with time-strided convolution (T-conv) used by default for connection.

**Overall Output.** The outputs of the two pathways are respectively subjected to global average pooling, and the resulting feature vectors are concatenated to get a final representation is of length 2304(2048+256).

### B.2. Slowfast audio encoder

As shown in Table B2, the structure of Slowfast audio encoder (Kazakos et al, 2021) is explicitly inspired by its video counterpart. Architecturally, they share several key similarities:

**Two-stream design with specialized pathways**. Both architectures employ a two-stream framework consisting of Slow and Fast pathways. The Slow stream is designed with high channel capacity to capture semantic information (frequency semantics for audio, spatial semantics for video), while the Fast stream operates at a finer temporal resolution with more temporal convolutions to focus on temporal patterns (temporal dynamics in audio, rapid motion changes in video).

**Multi-level lateral fusion**. Both integrate information across streams through multi-level lateral connections. Specifically, the Fast stream's output is processed to match the Slow stream's sampling rate, and the feature maps are then fused, enabling complementary information exchange between the two pathways to enhance overall representation capability.

**Residual network foundation**. Both streams in both architectures are variants of ResNet (e.g., ResNet-50), consisting of initial convolutional blocks with pooling layers followed by multiple residual stages, leveraging the residual learning mechanism to facilitate training of deep networks.

**Overall Output.** The final feature representation of the SlowFast audio encoder is generated through a consistent mechanism as the video encoder. The pooled feature vectors from both streams are concatenated, resulting in a final representation with a length of 2304(2048+256). Two feature vectors of identical dimension are then employed in the contrastive learning procedure.

## Appendix C. Implementation Details of the proposed MultiSoundGen

This study leverages Direct Preference Optimization (DPO), a reinforcement learning technique, to fine-tune our base model, thereby enhancing its audio generation quality and audio-visual alignment. The specific implementation details and hyperparameter configurations for the DPO fine-tuning are outlined below.

### C.1. Iterative Optimization Process

The entire optimization process consists of 5 iterations. The procedure for each iteration is as follows:

**Audio Generation**: The base model from the previous iteration is used to generate **5** candidate audio clips for each training video.

**Reward Assessment**: These 5 candidate audio clips, along with their corresponding original videos, are fed into our proposed SF-CAVP reward model for scoring.

**Preference Data Construction**: Based on the scores from the SF-CAVP reward model, the audio clip with the lowest score is designated as $\text{Audio}^{l}$ (inferior audio), while the original audio from the video is designated as $\text{Audio}^{w}$ (superior audio).

**Gradient Backpropagation**: The constructed preference pair ($\text{Audio}^{w}$, $\text{Audio}^{l}$) is used to compute the loss $L_{\text{AVP-RPO}}$, and gradients are backpropagated to update the model parameters.

**Model Update**: Each iteration comprises **1000** training steps, after which the resulting model becomes the base model for the subsequent iteration. This iterative process is repeated **5** times, culminating in the optimized MultiSoundGen model.

### C.2. DPO Training Parameters

The DPO training in this work was configured with specific hyperparameters to optimize model performance. Each iteration consisted of 1000 training steps, with a learning rate of $5.0\times10^{-6}$ and a weight decay of $1.0\times10^{-4}$. To ensure stable training, a linear warmup for the learning rate was applied over the first 100 steps, which constitutes approximately 10% of the total training steps. The learning rate schedule followed a cosine annealing policy. Gradient accumulation was set to 2, effectively increasing the batch size without additional memory overhead, and gradient norm was clipped at 1.0 to prevent exploding gradients.

### C.3. Computing Infrastructure and Inference Efficiency

Our experiments were conducted on a single server equipped with two NVIDIA H800 GPUs. The proposed method, MultiSoundGen, has a parameter count of 157M. The training process was managed using standard deep learning frameworks and libraries, including PyTorch, CUDA, and cuDNN, on a Linux operating system. In terms of inference efficiency, MultiSoundGen achieves a swift average generation time of 1.43s for an 8s audio clip, utilizing only 25 sampling steps. This performance is comparable to its base model, MMAudio, and significantly outperforms other competitive methods. Specifically, MultiSoundGen's generation speed is notably faster than V-AURA (33.63s) and Seeing&Hearing (30.18s), and also surpasses Foleycrafter (3.71s) and Frieren (2.68s). These results convincingly demonstrate that our method combines high-quality generation and audio-visual alignment with exceptional computational efficiency.

**Appendix D. Ablation of winner audio for preference creation**

To verify that using ground truth audio is more appropriate as the winner than the highest-scoring generated audio in preference creation, we conducted experiments with two strategies and compared their results, which are presented in Figure D1. Using the highest-scoring generated audio as the winner does not significantly improve model performance. It even deteriorates the IB-score progressively. In contrast, employing ground truth audio achieves better optimization results, with all metrics outperforming those of the former, with only Desync remaining on par.

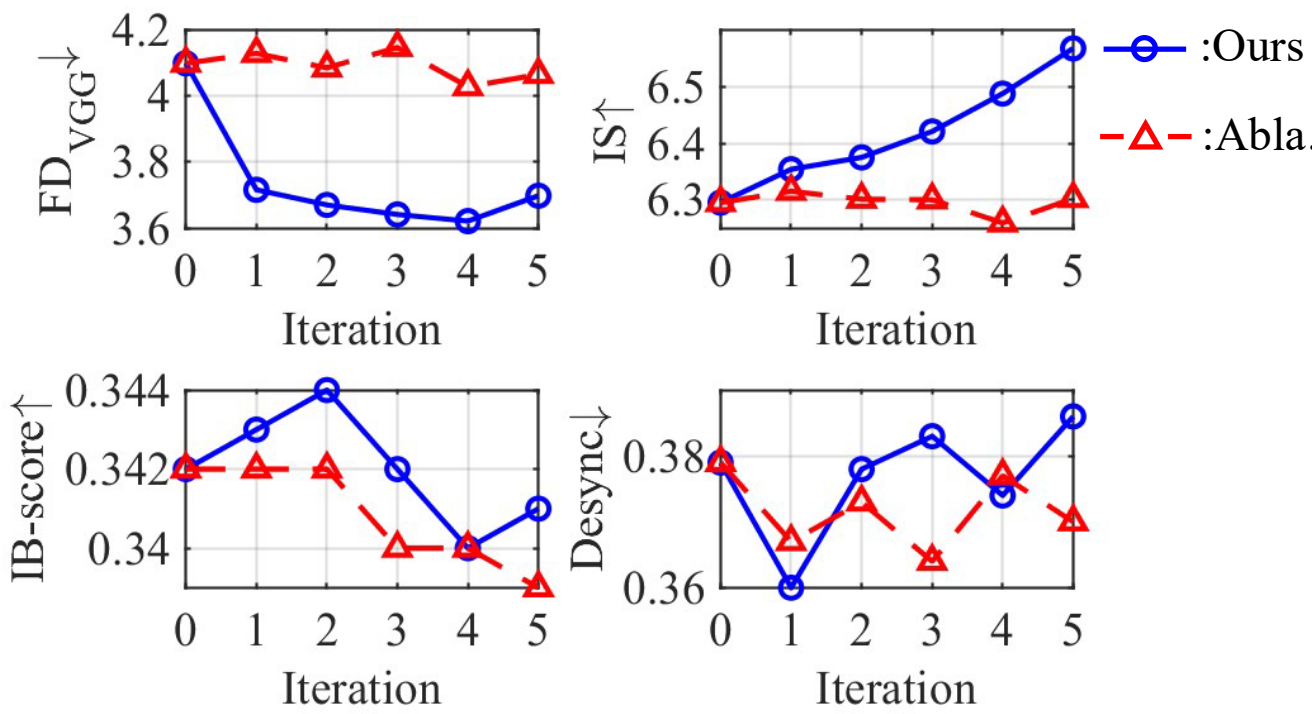

Figure D1. Comparison of winner selection strategies in preference creation. Using highest-scoring generated audio as winner (dashed line) fails to improve performance, while ground truth audio (solid line) yields better optimization, validating its effectiveness.